\documentclass[a4paper]{article}

\usepackage{INTERSPEECH2019}
\usepackage{url}
\usepackage{xcolor}
\usepackage{multirow}

\makeatletter
\g@addto@macro{\UrlBreaks}{%
\do\/\do\-\do\_}
\makeatother

\title{The Second DIHARD Diarization Challenge: Dataset, task, and baselines}
\name{Neville Ryant$^{1}$,
      Kenneth Church$^{2}$,
      Christopher Cieri$^{1}$,
      Alejandrina Cristia$^{3}$, 
      Jun Du$^{4}$, 
      Sriram Ganapathy$^{5}$,
      Mark Liberman$^{1}$
    }

\address{
  $^1$Linguistic Data Consortium, University of Pennsylvania, Philadelphia, PA, USA \\
  $^2$Baidu Research, Sunnyvale, CA, USA \\
  $^3$Laboratoire de Sciences Cognitives et de Psycholinguistique, D\'{e}pt d'\'{e}tudes cognitives, ENS, EHESS, CNRS, PSL University, Paris, France \\
  $^4$University of Science and Technology of China, Hefei, China \\ 
  $^5$Electrical Engineering Department, Indian Institute of Science, Bangalore, India}
\email{nryant@ldc.upenn.edu}

\begin{document}

\maketitle

\begin{abstract}
    This paper introduces the second DIHARD challenge, the second in a series of speaker diarization challenges intended to improve the robustness of diarization systems to variation in recording equipment, noise conditions, and conversational domain. The challenge comprises four tracks evaluating diarization performance under two input conditions (single channel vs. multi-channel) and two segmentation conditions (diarization from a reference speech segmentation vs. diarization from scratch). In order to prevent participants from overtuning to a particular combination of recording conditions and conversational domain, recordings are drawn from a variety of sources ranging from read audiobooks to meeting speech, to child language acquisition recordings, to dinner parties, to web video. We describe the task and metrics, challenge design, datasets, and baseline systems for speech enhancement, speech activity detection, and diarization.
\end{abstract}

\section{Introduction}

Speaker diarization, often referred to as ``who spoke when'', is the task of determining how many speakers are present in a conversation and correctly identifying all segments for each speaker. In addition to being an interesting technical challenge, it forms an important part of the pre-processing pipeline for speech-to-text and is essential for making objective measurements of turn-taking behavior. Early work in this area was driven by the NIST Rich Transcription (RT) evaluations \cite{fiscus2006rich}, which ran between 2002 and 2009. In addition to driving substantial performance improvements, especially for meeting speech, the RT evaluations introduced the diarization error rate (DER) metric, which remains the principal evaluation metric in this area. Since the RT evaluation series ended in 2009, diarization performance has continued to improve, though the lack of a common task has resulted in fragmentation with individual research groups focusing on different datasets or domains (e.g., conversational telephone speech \cite{sell2014speaker,zhu2016online,garcia2017speaker,wang2018speaker,zhang2019}, broadcast \cite{rouvier2013open,vinals2017domain}, or meeting \cite{yella2013improved,yella2014artificial}). At best, this has made comparing performance difficult, while at worst it may have engendered overfitting to individual domains/datasets resulting in systems that do not generalize. Moreover, the majority of this work has evaluated systems using a modified version of DER in which speech within 250 ms of reference boundaries and overlapped speech are excluded from scoring. As short segments such as backchannels and overlapping speech are both common in conversation, this may have resulted in an over-optimistic assessment of performance even within these domains\footnote{See, for instance, the release of IBM's diarization API in 2017. The feature worked well for simple cases, but when run by users on real inputs, the performance was found to be lacking, especially for overlaps, back-channels, and short turns.} \cite{milner2016segment}.

It is against this backdrop that the JSALT-2017 workshop \cite{ryant2018enhancement} and DIHARD challenges\footnote{\url{https://coml.lscp.ens.fr/dihard/index.html}} emerged. The DIHARD series of challenges introduce a new common task for diarization that is intended both to facilitate comparison of current and future systems through standardized data, tasks, and metrics and promote work on robust diarization systems; that is systems, that are able to accurately handle highly interactive and overlapping speech from a range of conversational domains, while being resilient to variation in recording equipment, recording environment, reverberation, ambient noise, number of speakers, and speaker demographics. As with the NIST RT evaluations, DER is adopted as the primary evaluation metric, but without use of collars or exclusion of overlapping speech. There are no constraints on training data, with participants allowed to use any combination of public/proprietary data for system development.

The initial DIHARD challenge (DIHARD I) \cite{dihard1_eval_plan} ran during the spring of 2018 and attracted registrations from 20 teams, of which 13 submitted systems. As expected, state-of-the-art systems performed poorly, with final DER on the evaluation set for the top systems ranging from 23.73\% \cite{sell2018diarization} when provided with reference speech activity detection (SAD) marks to 35.51\% \cite{diez2018but} when forced to perform diarization from scratch. These error rates rates are more than double the state-of-the-art for CALLHOME \cite{cieri2003switchboard} at the time \cite{garcia2017speaker,wang2018speaker}. For some domains, error rates for the best systems exceeded 49\% when using reference SAD and 75\% when performing diarization from scratch!

The second DIHARD Challenge (DIHARD II) \cite{dihard2_eval_plan}, like its predecessor, examines diarization system performance under two SAD conditions: diarization from a supplied reference SAD and diarization from scratch. As with DIHARD I, it includes a single channel input condition utilizing wideband speech sampled from 11 demanding domains, ranging from clean, nearfield recordings of read audiobooks to extremely noisy, highly interactive, farfield recordings of speech in restaurants to child language data recorded in the home using LENA vests. Unlike DIHARD I, it additionally offers a multichannel input condition requiring participants to perform diarization from farfield microphone arrays of dinner party speech drawn from the CHiME-5 corpus \cite{barker2018fifth}. For the first time, we also provide participants with baseline systems for speech enhancement, SAD, and diarization, as well as results obtained with these systems for all tracks.  

\section{Tracks}
\label{sec:tracks}
The challenge features two audio input conditions:
    \begin{itemize}
        \item {\bf Single channel}  --  Systems are provided with a single channel of audio for each recording. Depending on the recording source, this channel may be taken from a single distant microphone, a single channel from a distant microphone array, a mix of head-mounted or array microphones, or a mix of binaural microphones.
        \item {\bf Multichannel}  --  Each recording session contains output from one or more distant microphone arrays, each containing multiple channels. Participants are instructed to treat the arrays separately, producing one output per array. They are free to use as few or as many of the channels on each array as they wish to perform diarization.
    \end{itemize}
As system performance is strongly tied to the quality of the SAD component, we also include two SAD conditions:
    \begin{itemize}
        \item {\bf Reference SAD}  --  Systems are provided with a reference speech segmentation that is generated by merging speaker turns in the reference diarization.
        \item {\bf System SAD}  --  Systems are provided with just the raw audio input for each recording session and are responsible for producing their own speech segmentation.
    \end{itemize}
Together, this yields the following four evaluation tracks:
    \begin{itemize}
        \item {\bf Track 1}  --  single channel audio using reference SAD
        \item {\bf Track 2}  --  single channel audio using system SAD
        \item {\bf Track 3}  --  multichannel audio using reference SAD
        \item {\bf Track 4}  --  multichannel audio using system SAD
    \end{itemize}
All teams are required to register for at least one of track 1 or track 3.

\section{Performance Metrics}
\label{sec:metrics}
As in DIHARD I, the primary metric is DER \cite{fiscus2006rich}, which is the sum of missed speech, false alarm speech, and speaker misclassification error rates. Because systems are provided with the reference speech segmentation for tracks 1 and 3, for these tracks, it exclusively measures speaker misclassification error. This is the metric used to rank systems on the leaderboard.

For each system we also compute a secondary metric, Jaccard error rate (JER), which is newly developed for DIHARD II. JER is based on the Jaccard similarity index \cite{hamers1989similarity,real1996probabilistic},
a metric commonly used to evaluate the output of image segmentation systems, which is defined as the ratio between the sizes of the intersections and unions of two sets of segments. An optimal mapping between speakers in the reference diarization and speakers in the system diarization is determined and for each pair the Jaccard index of their segmentations is computed. JER is defined as 1 minus the average of these scores, expressed as a percentage. That is, it is the mean of Eq. \ref{eq:jer} across all reference speakers $ref$, where {\it TOTAL} is the duration of the union of reference and system speaker segments, {\it FA} is the total system speaker time not attributed to the reference speaker, and {\it MISS} is the total reference speaker time not attributed to the system speaker. It ranges from 0\% in the case where each reference speaker is paired with a system speaker with an identical segmentation to 100\% in the case where none of the system speakers overlap any of the reference speakers.
    \begin{equation}
        \textrm{JER}_{ref} = \frac{\textrm{FA} + \textrm{MISS}}{\textrm{TOTAL}}
        \label{eq:jer}
    \end{equation}

All metrics are computed using version 1.0.1 of the {\it dscore} tool\footnote{\url{https://github.com/nryant/dscore}} without the use of forgiveness collars and with scoring of overlapped speech.

\section{Datasets}
\label{sec:data}

\begin{table}[t]
    \centering
    \caption{Overview of DIHARD II datasets. For the CHiME-5 (multichannel) data, each Kinect is treated as a separate recording.}
    \begin{tabular}{c|c|c|c}
    \hline
    Input condition         & Set   & Duration (hours) & \# Recordings \\ 
    \hline
    \multirow{2}{*}{single channel} & dev   & 23.81       & 192    \\
    \cline{2-4}
                                    & eval  & 22.49       & 194    \\
    \hline
    \multirow{2}{*}{multichannel}  & dev   & 262.41      & 105    \\
    \cline{2-4}
                                    & eval  & 31.24       & 12     \\
    \hline
    \end{tabular}
    \label{tab:datasets}
\end{table}

\subsection{Overview}
The DIHARD II development and evaluation sets draw from a diverse set of sources exhibiting wide variation in recording equipment, recording environment, ambient noise, number of speakers, and speaker demographics. The single channel input condition (tracks 1 and 2) dataset is a superset of that used in DIHARD I, though 6 hours of additional material have been added to ensure that all domains are represented in both the development and evaluation set. Additionally, two domains where the DIHARD I annotation was deemed suspect (child language and web video) have been entirely resegmented. For the multichannel input condition (tracks 3 and 4) we use the multi-party dinner recordings originally collected for and exposed during the CHiME-5 challenge \cite{barker2018fifth}. The development and evaluation sets are summarized in Table \ref{tab:datasets}.

The development set includes reference diarization and speech segmentation and may be used for any purpose including system development or training. As with DIHARD I, there is no training set, with participants free to train their systems on any proprietary and/or public data. Both the development and evaluation sets will be submitted for publication via LDC at the end of the evaluation.

\subsection{Single channel data (tracks 1 and 2)}
\label{sec:singlechan}
The single channel input condition development and evaluation sets consist of selections of 5-10 minute duration samples drawn from 11 conversational domains, each including approximately 2 hours of audio. The full set of domains is described below with LDC Catalog numbers where appropriate. Unless otherwise specified, all speech is English, though not necessarily by native or even fluent speakers. All audio is distributed via LDC as 16 kHz, monochannel FLAC files.

\begin{itemize}
    \item {\bf audiobooks} -- amateur recordings of public domain English works drawn from LibriVox; care was taken to avoid overlap with LibriSpeech \cite{panayotov2015librispeech} (unpublished)
    \item {\bf broadcast interview} -- student produced interviews with newsmakers of the day taken from a late 1970s college radio show; recorded on open reel tapes before being digitized and contributed to LDC (unpublished)
    \item {\bf child language} -- day-long recordings of 6-18 month old vocalizations collected at home by University of Rochester researchers for the SEEDLingS corpus \cite{bergelson2016bergelson}
    \item {\bf clinical} -- interviews with 12-16 year old children intended to determine whether or not they fit the clinical diagnosis for autism; all recordings conducted at the Center for Autism Research (CAR) of the Children’s Hospital of Philadelphia (CHOP) using a mixture of cameras and ceiling mounted microphones (unpublished)
    \item {\bf courtroom} -- oral arguments from the 2001 term of the U.S. Supreme Court that were  digitized for the OYEZ project; recordings are summed from individual table-mounted microphones, one per speaker (unpublished)
    \item {\bf map task} -- recordings of map tasks in which one participant, the leader, describes a route drawn on a map to the other participant, the follower, who attempts to draw the same route on a copy of the map lacking the route and optionally lacking some landmarks; audio was recorded via close-talking microphones under quiet conditions (previously released as LDC96S38)
    \item {\bf meeting} -- meetings with between 3 and 7 participants, each recorded with a variety of close-talking and distant microphones, from which a single, centrally located distant microphone was selected; the development set draws from the NIST Spring 2004 Rich Transcription Evaluation (LDC2007S11 and LDC2007S12) while the evaluation set draws from previously upublished recordings conducted for the DARPA Robust Omnipresent Automatic Recognition (ROAR) project at LDC in 2001
    \item {\bf restaurant} -- $\approx$1 hour sessions involving 3-6 diners recorded on a binaural microphone worn by one participant in restaurants with varying room acoustics and noise levels; inspired by the NSF Hearables Challenge and extended by LDC for DIHARD (unpublished)
    \item {\bf sociolinguistic field recordings} -- sociolinguistic interviews recorded under field conditions during the 1960s and 1970s; recorded under diverse locations and conditions with subjects ranging from 15 to 81 years of age and representing diverse ethnicities, backgrounds, and dialects of world English; the development set draws from SLX (LDC2003T15) and the evaluation set from DASS (LDC2012S03 \& LDC2016S05)
    \item {\bf sociolinguistic lab recordings} -- sociolinguistic interviews recorded as part of MIXER6 (LDC2013S03) under quiet conditions in a controlled environment; sessions were recorded with a variety of close-talking and distant microphones from which a single, centrally located distant microphone was selected
    \item {\bf web video} -- English and Mandarin amateur videos collected from online sharing sites (e.g., YouTube and Vimeo) as part of the Video Annotation for Speech Technologies (VAST) \cite{tracey2018vast} collection (mostly unpublished)
    \end{itemize}

\subsection{Multichannel data (tracks 3 and 4)}
\label{sec:multichan}
The multichannel input condition development and evaluation sets are drawn from the CHiME-5 dinner party corpus \cite{barker2018fifth}, a corpus of conversational speech collected during dinner parties held in real homes. The development set combines the CHiME-5 training and development sets and encompasses 45 hours of dinner parties from 18 homes. The evaluation set is identical to the CHiME-5 evaluation set and consists of 5 hours of dinner parties from 2 homes. Each party was recorded using 6 Microsoft Kinect devices (4 channel linear arrays) distributed throughout the home in such a way that the conversation was always present on each array. Due to a combination of clock drift and random frame dropping, the Kinects within each recording session exhibit massive desynchronization, both with each other and with the binaural recording devices worn by participants. For this reason, each Kinect device is treated separately with the resulting development and evaluation sets having durations of 262.4 hours and 31.2 hours respectively. All audio is distributed via the University of Sheffield as 16 kHz WAV files.

\subsection{Processing}
A limited number of recordings contained regions carrying personal identifying information (PII), which were removed prior to publication. For the {\it clinical} and {\it restaurant} domains, this was done at LDC by low-pass filtering using a 10th order Butterworth filter with a passband of 0 to 400 Hz. To avoid abrupt transitions in the resulting waveform, the effect of the filter was gradually faded in and out at the beginning and end of the regions using a ramp of 40 ms. In the case of the {\it sociolinguistic field recordings} domain and the CHiME-5 data, PII was removed by the original creators of the corpora. In the former case, PII was replaced by tones of matched duration, while in the latter case it was zeroed out. PII containing regions are ignored during scoring.

\subsection{Annotation}
Reference segmentation and speaker labeling was produced by annotators at LDC using a tool equipped with playback, waveform and spectrogram display. Annotators were instructed to split on pauses $>$ 200 ms, where a pause was defined as any stretch of time during which the speaker was not producing vocalization (e.g., backchannels, filled pauses, singing, speech errors and disfluencies, infant babbling or vocalizations, laughter, coughs, breaths, lipsmacks, and humming) of any kind. Boundaries were placed within 10 ms of the true boundary, taking care not to truncate sounds at edges of words (e.g., utterance-final fricatives). Where individual close talking microphones were available for speakers, annotation was performed separately for each speaker using their individual microphone. Due to time constraints, this manual segmentation process could not be implemented for the multichannel development data; for this data, segmentation was taken from the turn boundaries established during the original CHiME-5 transcription. 

An additional post-processing step was necessary for the CHiME-5 annotation to correct for the lack of synchronization between binaural recording devices and Kinects. For each Kinect, the lag between that array and the binaural recording devices was estimated at regular intervals using normalized cross-correlation. The speech boundaries etablished by annotation on the binaural devices were then corrected for each Kinect using these estimated lags.

\section{Baseline system}
\label{sec:baseline}
\subsection{Speech enhancement}
For speech enhancement we use a densely-connected LSTM architecture \cite{gao2018densely,sun2018speaker,sun2018novel} trained to predict the ideal ratio masks (IRM) \cite{srinivasan2006binary} of speech from log-power spectra (LPS) features. The model is trained via progressive multi-target learning \cite{gao2018densely,sun2017multiple} using 400 hours of noisy speech produced by corrupting clean utterances from WSJ0 \cite{wsj0} and a 50 hour Chinese speech corpus from the 863 Program \cite{qian2004introduction}. Utterances were corrupted using 115 noise types \cite{gao2018densely} at 3 SNR levels (-5dB, 0dB, and 5dB). The trained models as well as scripts for applying them, are distributed through GitHub\footnote{\url{https://github.com/staplesinLA/denoising_DIHARD18}\label{foot:denoise}}.

\subsection{Beamforming}
For the multichannel tracks, we use weighted delay-and-sum beamforming as implemented in BeamformIt \cite{anguera2007acoustic}. Beamforming is applied independently for each Kinect  in each session using all four channels following the CHiME-5 recipe \cite{barker2018fifth}.

\subsection{Speech activity detection}
The baselines for tracks 2 and 4 use WebRTC's\footnote{\url{https://webrtc.org/}} SAD as implemented in the {\it py-webrtc} Python package\footnote{\url{https://github.com/wiseman/py-webrtcvad}}. Scripts for performing SAD using the same settings used to obtain the baseline results are distributed through GitHub\textsuperscript{\ref{foot:denoise}}.

\subsection{Diarization}
The diarization baseline is based on the previously published Kaldi \cite{povey2011kaldi} recipe\footnote{\url{https://github.com/kaldi-asr/kaldi/tree/master/egs/dihard_2018/v2}} for JHU's submission to DIHARD I \cite{sell2018diarization}. At a high level, the system performs diarization by dividing each recording into short overlapping segments, extracting x-vectors \cite{snyder2016deep,snyder2018x}, scoring with probabilistic linear discriminant analysis (PLDA) \cite{prince2007probabilistic}, and clustering using agglomerative hierarchical clustering (AHC) \cite{han2008strategies}. In contrast to the original JHU system, we omit the Variational Bayes resegmentation step \cite{diez2018speaker}. The trained models are distributed through GitHub\footnote{\url{https://github.com/iiscleap/DIHARD_2019_baseline_alltracks}}.

The x-vector extractor configuration is identical to that used in previous speaker recognition and diarization systems \cite{snyder2018x,sell2018diarization} with two exceptions: i) $30$ dimensional mel frequency cepstral coefficient (MFCC) features are used instead of mel filterbank features; ii) the embedding layer uses 512 dimensions. MFCCs are extracted every $10$ ms using a $25$ ms window and mean-normalized using a 3 second sliding window. For training we use a combination of VoxCeleb 1 and VoxCeleb 2 \cite{nagrani2017voxceleb,chung2018voxceleb2} augmented with additive noise and reverberation according to the recipe from \cite{snyder2016deep}. Segments under 4 seconds duration are discarded, resulting in a training set with 7,323 speakers. Reverberation is added by convolution with room responses from the RIR dataset \cite{ko2017study}, while additive noises are drawn from the MUSAN dataset \cite{snyder2015musan}. At test time, x-vectors are extracted from 1.5 second segments with 0.75 second overlap.

Following extraction, x-vectors are pre-processed to perform domain adaptation to the DIHARD II dataset. This is done by normalizing with a global mean and whitening transform learned from the DIHARD II development set. The whitened x-vectors are then length normalized \cite{garcia2011analysis} and used to train a Gaussian PLDA model \cite{prince2007probabilistic} using a subset of VoxCeleb consisting of segments of at least $3$ seconds duration. Following PLDA scoring, clustering is performed using AHC with the threshold set by minimizing DER on the development data.

\begin{table}[t]
    \centering
    \caption{Baseline performance (measured by DER and JER) on dev and eval sets for all tracks. The Enh. column indicates whether or not speech enhancement was applied prior to SAD.}
    \begin{tabular}{c||c|c|c|c|c}
    \hline
         Track  & Enh.  &  \multicolumn{2}{c|}{DER (\%)} & \multicolumn{2}{c}{JER (\%)} \\ 
                &            & Dev       & Eval     & Dev       & Eval \\ 
         \hline
         Track 1 & no        & 23.70     & 25.99     & 56.20    & 59.51 \\
         \hline
         Track 2 & no        & 46.33     & 50.12     & 69.26    & 72.1 \\
         \hline
         Track 2 & yes       & 38.26     & 40.86     & 62.59    & 66.60 \\
         \hline 
         Track 3 & no        & 59.73     & 50.85     & 68.00    & 65.91  \\
         \hline
         Track 4 & no        & 87.55     & 83.41     & 88.08    & 85.12  \\
         \hline
         Track 4 & yes       & 82.49     & 77.34      & 83.6    & 80.42  \\
         \hline
    \end{tabular}
    \label{tab:baselineresults}
\end{table}

\subsection{Baseline results}
DER and JER of the baseline system on both the development and evaluation sets for each track are presented in Table~\ref{tab:baselineresults}. The speech enhancement module is used only for tracks 2 and 4 as a pre-processing front-end for the SAD pipeline as the diarization system did not show improvements using the enhanced audio. The scores obtained by the challenge baseline are quite high, with track 1 DER roughly in line with the performance of the best DIHARD I systems \cite{sell2018diarization,diez2018but,sun2018speaker} and track 2 DER 5\% higher than for DIHARD I (15\% without enhancement), which we suspect reflects a combination of superior SAD components in those systems and the more careful segmentation for the child language and web video domains in DIHARD II. Error rates are noticeably higher for tracks 3 and 4, reaching 50.85\% and 77.34\% respectively, though, again, these rates are roughly in line with those observed for the best DIHARD I systems on the two most difficult domains in that challenge: {\it restaurant} and {\it child language}.

\section{Conclusion}
The field of speaker diarization has changed drastically in the two short years we have been running this challenge. In the lead up to DIHARD I, the research community was fragmented and most research concentrated on relatively easy datasets using forgiving evaluation metrics. This both made comparison of systems difficult and led some to believe that diarization was relatively solved and uninteresting. However, we were pleased by the response to DIHARD I, both during the evaluation and after, demonstrating that there is interest in robust diarization. This renewed energy is on display in DIHARD II, which attracted 48 registered teams from 17 countries, more than doubling the number of teams registered for DIHARD I. It is also evident in the recent announcement of the Fearless Steps challenge, which includes diarization among its tasks. We hope that this year's contributions lead to marked progress toward the goal of truly robust diarization.

\section{Acknowledgements}
We would like to thank Harshah Vardhan MA, Prachi Singh, and Lei Sun for their help in preparing the baseline sytems and results. We would also like to acknowledge the generous support of Agence Nationale de la Recherche (ANR-16-DATA-0004 ACLEW, ANR-14-CE30-0003 MechELex, ANR-17-EURE-0017), the J. S. McDonnell Foundation, and the Linguistic Data Consortium as well as the CHiME-5 challenge for allowing us use of their data.

\clearpage
\bibliographystyle{IEEEtran}
\bibliography{refs}

\end{document}